# TeraHz tuning of whispering gallery modes in a PDMS, stand-alone, stretchable microsphere

Ramgopal Madugani,[1,2,*] Yong Yang (杨勇),[1,2] Jonathan M Ward,[1,2] John Daniel Riordan,[1] Sara Coppola,[3] Veronica Vespini,[3] Simonetta Grilli,[3] Andrea Finizio,[3] Pietro Ferraro,[3] and Síle Nic Chormaic[1,2,4]

[1]*Physics Department, University College Cork, Cork, Ireland*
[2]*Photonics Centre, Tyndall National Institute, Prospect Row, Cork, Ireland*
[3]*CNR Istituto Nazionale di Ottica—sezione di Napoli, via Campi Flegrei 34, 80078 Pozzuoli (Napoli), Italy*
[4]*Light-Matter Interactions Unit, OIST Graduate University, 1919-1 Tancha, Onna-son, Kunigami, Okinawa 904-0495, Japan*
*\*Corresponding author: ramgopal.madugani@tyndall.ie*

We report on tuning the optical whispering gallery modes in a poly dimethyl siloxane-based (PDMS) microsphere resonator by more than a THz. The PDMS microsphere system consists of a solid spherical resonator directly formed with double stems on either side. The stems act like tie-rods for simple mechanical stretching of the microresonator over tens of microns, resulting in tuning of the whispering gallery modes by one free spectral range. Further investigations demonstrate that the whispering gallery mode shift has a higher sensitivity (0.13 nm/µN) to an applied force when the resonator is in its maximally stretched state compared to its relaxed state.

Tuning the whispering gallery modes (WGMs) of an optical microcavity is a key requirement for many applications, for example, biological sensing [1], chemical sensing [2] or quantum optics [3,4]. While tuning is a desirable feature of optical microcavities it can be difficult to implement in a compact and integrated way. The main tuning techniques are temperature tuning via both internal and external means [5-8], stress/strain tuning [3,9-12], optomechanical (dispersive) tuning [13,14] and chemical etching [15]. Magnetorheological [16] and electrostriction effects [17] have also been reported. Each of these methods has its own advantages and disadvantages and the implementation of any particular method will depend on the size, material and geometry of the microcavity under investigation. For example, chemical etching is used to change the cavity size so that resonance occurs, but is not a real-time tuning solution. While in [17] water-filling in hollow poly dimethyl siloxane (PDMS) spheres was required to achieve strain-sensitivity to electric fields. Tuning of the microcavity over one free spectral range (FSR) is desirable; however, the FSR increases as the size of the cavity decreases, thereby making it even more difficult to achieve the desired tuning effect.

Stress/strain tuning is usually realized by applying a mechanical force (compressive or stretching) to the microcavity. The applied force deforms the cavity shape and changes the refractive index by the elasto-optic effect [3,9,11]. The resonance wavelengths of the WGMs in polymeric microspheres, for example, are shown to have high sensitivity to mechanical forces [9]. The sensitivity of a microcavity to stress/strain tuning can be increased if the cavity is hollow [9,11,18]. A WGM tuning range of around 5 nm at a wavelength of 1550 nm has been observed for a hollow bottle resonator under axial tension [18]. Alternatively, microbubble cavities can be filled with a fluid, which can then be compressed to achieve stress/strain tuning [11]. Fine tuning of WGMs in a double stem, silica microsphere was initially reported in [3] and more recently, 500 GHz strain tuning in a double-stem, silica sphere was demonstrated [12].

Here, we report on tuning WGMs over 15 nm (1.9 THz) in a stand-alone, stretchable, polymeric, spherical microresonator. The microsphere is directly shaped during the fabrication process, with double stems on opposite sides that can be used as tie-rods. Such tie-rods facilitate easy stretching of the microsphere without the need for any interaction or contact of mechanical tools with the resonator, hence the term "stand-alone". The microsphere is formed using a liquid electro-drawing 3D-lithography approach [19] based on liquid instabilities induced by electrohydrodynamic (EHD) pressure [20,21].

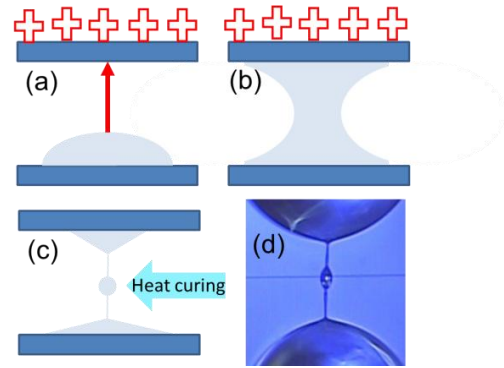

Fig. 1. (Color online) Schematic of microsphere formation. (a) PDMS droplet on a glass substrate below a charged lithium niobate (LN) plate. The arrow indicates the force drawing the polymer upward. (b) The liquid forms a bridge between the glass substrate and the LN. (c) The bridge thins and the flow of liquid creates a liquid bead. The PDMS is rapidly thermally cured creating a solid sphere. (d) An image of a 37 µm solid oblate PDMS sphere on a string of length, $L \sim 237$ µm, the stem radii are $\sim 6$ µm. The coupling tapered optical fiber is also visible.

In this case, the pyro-EHD (PEHD) method is used [19,21] (see Fig. 1). A liquid PDMS droplet is placed on a glass substrate. A lithium niobate (LN) crystal is positioned over the droplet and the opposite side of the crystal is heated. This creates a positive charge on the surface of the crystal that draws the liquid upwards, forming a liquid PDMS bridge between the glass and the crystal. Once the bridge is formed it thins and instabilities cause liquid nanodroplets to form on it, spontaneously (i.e. beads-on-a-string). Rapid curing by a thermal stimulus results in solid PDMS microspheres on the string (see Fig. 1(d)). For the work presented here this process was performed between two glass capillaries so the sphere can be easily handled during experiments.

Usually, the shape of PDMS material, which has a low Young's modulus compared to glass and silica, can be changed by means of an applied force [9-12]. A 37 μm diameter PDMS microsphere, with a FSR of 14.7 nm determined by 1.5-1.55 μm scanning of the laser, is used to study the WGM shift due to stretching.

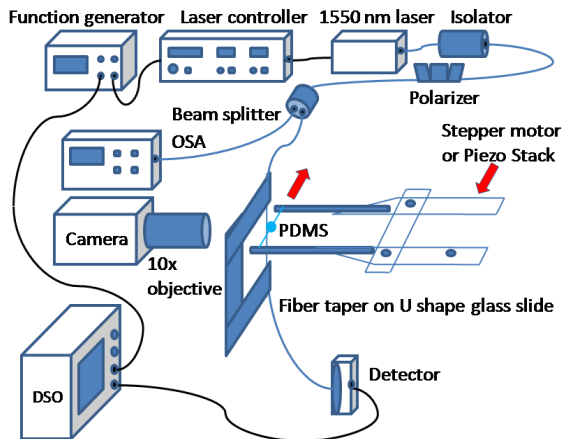

Fig. 2. (Color online) Experimental setup. The transmitted WGM spectra are detected by a photodiode and recorded on a digital storage oscilloscope (DSO). A stepper-motor-driven translation stage, with a 1 μm resolution, is used to push the mount and stretch the PDMS sphere. The arrows indicate the direction of the motion during stretching. OSA: optical spectrum analyzer.

The microsphere was directly fabricated onto a mechanical fiberglass frame composed of two arms to permit stretching. One arm of the frame was held in a fixed position while the other arm was free to pivot around a fixed point, see Fig. 2. The free end of the movable arm was pushed or pulled by a stepper motor driven translation stage (step size 2 μm). This gives an angular displacement to the associated capillary and, hence, a continuous stretching of the PDMS microsphere was possible. For submicron stretching a piezo stack with a 4 μm range was included between the frame and the stepper motor arm. Light was coupled into the microsphere via a tapered optical fiber with a diameter ~ 1 μm. The tapered fiber was in contact with the sphere throughout the experiment and some slack was allowed in the fiber.

To determine the tuning range, a typical WGM was selected. The observed Q factor of the sphere was ~$10^4$. The tension of the sphere was more or less arbitrary, though it was slightly pretensioned. When the sphere was stretched the WGM was blue-shifted and this was observed as a change in the position of the WGM on a 30 GHz laser scan window on the oscilloscope. The laser wavelength was adjusted to bring the WGM back to its initial position in the scan window, and the wavelength shift was noted using an optical spectrum analyzer (OSA).

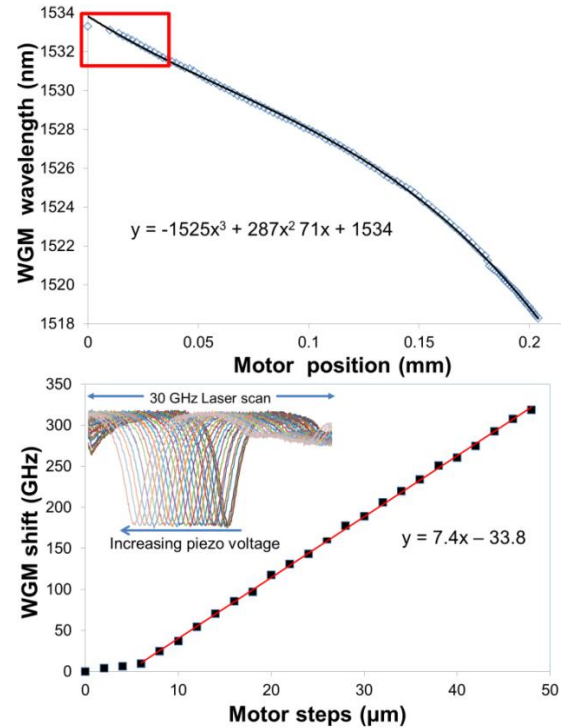

Fig. 3. (Color online) WGM shift for increased stretching. Top: WGM wavelength shift versus motor position. The total shift is over 1 FSR (~15 nm). The solid line is a third order polynomial fit. The region highlighted by the box is shown in the lower plot. Bottom: The WGM frequency shift versus motor position. The inset shows the fine tuning of the WGM using the piezo actuator. The piezo is rated for 4.0 ± 1.5 μm displacement for 150 V.

It was possible to follow a single WGM in this way while tuning over a large blue shift (~10 nm), although sometimes the coupling of the mode became too weak and another mode had to be selected. Care was taken to ensure modes with similar shift rates were used. The summation of these mode shifts is plotted in Fig. 3 (top) for a 200 μm movement of the motorized translation stage. The actual elongation, $\Delta L$, of the PDMS string was approximately 106 μm, determined from images taken with the CCD camera. The total recorded blue shift was larger than 15 nm. Between 10 μm and 130 μm the shift rate is linear at 0.059 nm (7.5 GHz)/μm, after 130 μm the shift rate increases nonlinearly to 0.14 nm (18 GHz)/μm. The area in the box in the top plot in Fig. 3 is shown again in the bottom plot of Fig. 3. In this region the shift rate is quite low due to the low tension on the sphere. Thereafter the tension reaches a threshold and the shifting rate increases sharply to a slope of 7.5 GHz/μm. The inset in Fig. 3 is an image of the WGM shift achieved using the piezo to stretch the sphere.

The nonlinear shift rate is shown again in Fig. 4. Each data point in Fig. 4 corresponds to a fixed displacement of the piezo plotted for different motor positions. Over the elongation of the sphere the shifting rate (for a 0-20 V change in the piezo voltage) increased from 6.5 GHz to 23

GHz. The stepper motor position was not shifted more than 350 μm to avoid risk of damage. The actual elongation was 185 μm (Δ$L/L$ = 0.64)

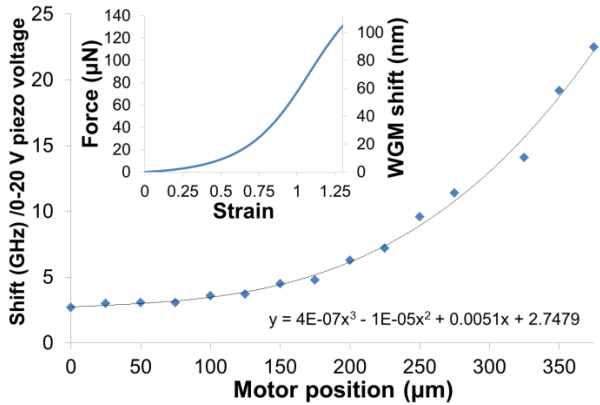

Fig. 4. (Color online) Sensitivity measurements showing the WGM shift for fixed piezo voltage at different motor positions. The solid line is a third order polynomial fit. Inset: theoretical plot showing the WGM shift dependence on applied force.

For each motor position the WGMs were observed to shift without applying any piezo voltage. Therefore, for each data point in Fig. 4 some time was needed for the drift to stop before applying the piezo voltage. This drift increased as the sphere was tensioned more. The largest drift occurred for about ten minutes. The origin of this behavior is unclear, but could arise from the nonlinear mechanical properties of PDMS [22,23]. The inset in Fig. 4 is a theoretical plot of the applied force and wavelength shift for a given strain. The WGM shift was calculated from equations in [9,24]. The parameters used were an elasto-optical constant = $-1.75 \times 10^{-10}$ m$^2$/N, shear modulus = $3.3 \times 10^5$ Pa, Young's modulus = $10^6$ Pa, Poisson's ratio = 0.5, PDMS microsphere radius = 19 μm, stem radius = 6 μm, refractive index = 1.4, and laser wavelength = 1550 nm. The stress, σ, and force were determined from experimental results in [22] and the data was fitted using the equation σ = $\mathrm{Arctan}(a(e^{b\varepsilon} - 1)10^6$, where $a = 0.01892$, $b = 3.675$ and ε is the strain (Δ$L/L$). The calculated force is only a rough estimate of the applied force. The model assumes the contact area between the sphere and stems does not change. It also does not take into account the relationship between the applied force and the changing shape of the sphere. Nonetheless it implies a force sensitivity of 0.13 nm/μN and, even with the low Q, the force resolution [9] is = $1.7 \times 10^{-7}$ N.

In summary, a large WGM tuning range in a stand-alone PDMS sphere on a string was measured and a nonlinear threshold was observed. Although some drift of the WGMs was seen under high tension, the WGM shift sensitivity was quite high in the linear region. The study of the WGMs could be used to infer mechanical properties of the PDMS. Moreover potential applications exist in chemical and biological sensing if the microspheres are appropriately functionalized.

Jonathan Ward acknowledges funding from the Irish Research Council for Science, Engineering and Technology (IRCSET) under the INSPIRE fellowship. This work is funded by OIST Graduate University, Japan.